1# UAV-aided urban target tracking system based on edge computing

Yajun Liu, Congxu Zhu, Xiaoheng Deng, Peiyuan Guan, Zhiwen Wan, JieLuo, Enlu Liu and Honggang Zhang

*Abstract*—Target tracking is an important issue of social security. In order to track a target, traditionally a large amount of surveillance video data need to be uploaded into the cloud for processing and analysis, which put stremendous bandwidth pressure on communication links in access networks and core networks. At the same time, the long delay in wide area network is very likely to cause a tracking system to lose its target. Often, unmanned aerial vehicle (UAV) has been adopted for target tracking due to its flexibility, but its limited flight time due to battery constraint and the blocking by various obstacles in the field pose two major challenges to its target tracking task, which also very likely results in the loss of target. A novel target tracking model that coordinates the tracking by UAV and ground nodes in an edge computing environment is proposed in this study. The model can effectively reduce the communication cost and the long delay of the traditional surveillance camera system that relies on cloud computing, and it can improve the probability of finding a target again after an UAV loses the tracing of that target. It has been demonstrated that the proposed system achieved a significantly better performance in terms of low latency, high reliability, and optimal quality of experience (QoE).

*Index Terms*—Target tracking, UAV, edge computing, low latency, high reliability, QoE.

## I. INTRODUCTION

Target tracking is an important issue in the area of public safety. It can be applied in a wide range of scenarios, such as the suspicious criminal tracking, and illegal vehicle tracking. The objective of target tracking in this study was to deal with the problem of the loss of tracking target via associating with air and ground nodes. However, the association can be especially difficult when the objects are fast moving so that ground cameras cannot catch up with. Another situation that increases the complexity of the problem is that a tracked object is blocked by obstacles when it changes orientation over time. In addition, target tracking can be a time consuming process, as it needs to use object recognition techniques for tracking, which is also quite challenging. As for target tracking in a city, target tracking application often require near real-time or real-time tracking. However, the time-consuming computation and complex terrain causes long delay to target tracking with those traditional tracking systems, such as ground surveillance cameras system, which cannot meet the real-time requirement of tracking applications.

Unmanned Aerial Vehicle (UAV) is usually used for missions that are too dangerous or difficult for humans. With the development of wireless communication techniques, the usage of UAV is rapidly expanding from military to commercial, recreational, agricultural, and other applications, such as product deliveries and surveillance. Due to its flexibility and wireless communication capability, UAVs have the following advantages in target tracking: 1) It can get real-time video of the target. 2) It can communication with ground-based controllers in a real-time manner. However, its flight time duration and target blocking by obstacles are still the two major challenges of target tracking for UAVs. Naser Hossein Motlagh et al. [1] introduced a UAV-based crowd surveillance use case to ensure a flight time long enough for UAVs by offloading the processing of video data to an edge cloud, i.e., mobile edge computing (MEC). Nevertheless, the target blocking by obstacles is still a problem so that a UAV cannot track a target. Especially, for urban target tracking, the complexity of terrain makes it difficult for a UAV to recover from losing a target by itself, which cannot be tolerated for a reliable tracking system. For example, when a suspicious vehicle enters an underground garage in order to avoid UAV tracking, it can escape from any exit of the garage. In that case, a UAV will lose its target due to a wide range of invisible areas and may not find the target again in a short period of time.

To address the aforementioned challenges in UAV target tracking, a ground surveillance camera system can be employed to form a cooperative air-ground network and to assist a UAV when the UAV cannot keep tracking due to the blocking of target by obstacles, and in the meantime, the computation tasks of target tracking can be offloaded to a ground computing cluster, which consists of mobile phones, laptops and edge servers. The three-layer air-ground network of this study was motivated by the following scenarios. Firstly, a UAV collects the video about a target in a real-time manner, and then it transmits the data to the ground computing cluster to analyze and recognize the target, and then the result is fed back to the UAV for further tracking guidance. In this case, an efficient stream processing strategy is needed, considering the movement of the tracking task, e.g., the Mobile Resource Aware (MRA) stream processing scheduling [7]. Secondly, when the UAV is about to lose track of the target, it needs to ask the ground surveillance cameras system to continue the tracking, and collect the surrounding information and transmit them to the nearby computing cluster for computation and recognition. Once the target is found again by a ground camera, the UAV can fly to the relevant location to continue tracking according to the geographic location information of the ground camera. Due to the limited power supply of the UAV and the real time requirement of the tracking system, an effective and fast activation strategy whose objectives are to minimize the activation time and the number of surveillance cameras, is therefore proposed; i.e., Efficient Activation Strategy. Thirdly, in the worst case scenario, in which the target is completely lost, all video data from nearby cameras need to be obtained right from the place where the target was lost, in order to find the target again. In order to quickly find the target again, the large amount of the computation tasks can be offloaded into a nearby ground computing cluster. For the case where there are computing-rich edge servers in the nearby cluster, a QoE-driven computation offloading strategy, which tackles the offloading scheduling problem via jointly allocating communication and computation resources with the consideration of the QoE (quality of experience) of the UAV is proposed [8]. When there are many mobile phones or other computing terminals and a few edge servers in the cluster, a Cloudlet Assisted Cooperative Task Assignment (CACAT) Strategy [9] that organizes edge nodes that are geographically close to a surveillance camera into a cluster is therefore proposed; to collaboratively work on the tracking task to minimize its total cost which is a weighted combination of latency and the costs incurred in working on the tracking task.

This article is organized in the following fashion. The related work about incorporating UAVs into ground networks is described in Section II. In Section III, the three-layer network for target tracking is introduced and in Section IV, the current results for each control strategy is discussed. Conclusions are finally given about the air-ground network for target tracking in Section V.

## II. RELATED WORK

Incorporating UAVs into ground networks, such as vehicular *ad hoc* networks (VANETs), radio access network, and Intelligent Transport Systems (ITS), have more and more applications. One of the main applications is disaster relief and management, where UAV-aided wireless communication provides reliable wireless connectivity for devices without infrastructure coverage due to damaged and dysfunctional by natural disasters. Sherman et al. [2] proposed an aerial-ground cooperative vehicular networking architecture for searching and rescuing after an earthquake, where UAVs are used to collect the road condition and deliver information to VANETs. Naser Hossein Motlagh et al. [1] introduced a crowd surveillance use case based on UAV and studied the offloading of video data processing to mobile edge computing (MEC) node compared to the local processing of video data onboard UAVs. They demonstrated the efficiency of the MEC-based offloading approach in saving the energy of UAVs, reducing the delay of recognition of suspicious persons by a test bed. Mohsen Guizani et al. [3] discussed the use of UAVs with broadband wireless technologies, which can augment the operation of public safety networks. Hamid Menouar et al. [4] investigated the potential and challenges for incorporating UAVs into LTE base stations to enable ITS for smart city, where UAVs are used as flying roadside units to capture video recordings of the incident scene and then relay them to a mobility services center.

But there are some challenges for UAVs joint with ground networks. Arvind Merwaday et al. [5] tackled the interference between UAVs, which are used as unmanned aerial base stations (UABs) in the heterogeneous network, in order to improve the throughput coverage and the $5^{th}$ percentile spectral efficiency of the network during natural disasters. Mohammad Mehedi Hassan et al. [6] tackled the problem of low complexity target tracking using flying robots.

However, all the above work for disasters relief is different from the target tracking, where the location of the disaster is determined while the tracked target is moving and uncertain. Moreover, UAV cannot catch up with a mobile target in a real time manner but can capture a video recording of the disaster. Therefore, proper mechanisms should be designed to enhance reliability and real-time of a target tracking system.

To deal with these issues in a target tracking system, a target tracking architecture for UAV combined with ground surveillance cameras and computing equipment is therefore proposed to deal with possible situations in a near real-time manner such as the loss of the UAV target.

## III. TARGET TRACKING ARCHITECTURE FOR UAV COMBINED WITH GROUND CAMERA AND COMPUTING EQUIPMENT

In a traditional UAV target tracking application applied to the smart city, the UAV is likely to lose the target due to the complexity of the surface topography. On the one hand, the suspected target is blocked by obstructions or the target is moving too fast to catch up with. On the other hand, the endurance of UAV is not enough, especially aerial UAV, which is generally not more than half an hour, and its computing power is also not enough to support real-time tracking task computation.

Compared with the UAV target tracking network, the target tracking network for UAV combined with ground surveillance cameras and ground nodes in an edge environment is more efficient. Due to wide geographical distribution and adequate endurance of ground surveillance cameras, they can be activated by UAV for relay tracking when the target is about to be lost in UAV's vision. However, the ground surveillance cameras have weak computing power so that it is difficult to support real-time target tracking, it is necessary for ground cameras to accomplish videos analysis and target recognition with ground edge computing cluster. For example, when the target is temporarily disappearing in the UAV's field of vision due to the barrier occlusion, the UAV can activate some nearby surveillance cameras to relay tracking according to their locations. Then, the ground surveillance cameras and nearby computing cluster start tracking after receiving suspicious target pictures and the recognition tracking model. Finally, the UAV decides the flight path or continues to activate other cameras based on feedbacks from the computing cluster. Therefore, the problem about poor computing capability of ground cameras can be dealt with and the performance of the tracking system can be improved when the UAV cannot catch up with the target.

Based on the above analysis of the cooperative network and the development of edge computing, a new architecture is therefore proposed. In this architecture, the UAV combines ground cameras and computing devices to build an aerial-ground cooperative surveillance system. Figure 1 depicts the cooperative surveillance system of the UAV when combined with ground cameras and computing devices.It is composed of a UAV, used as a collector and correspondent, a ground surveillance cameras network and a ground computing cluster. In this section, scenarios of the UAV-aided urban target tracking system, system components and the three-layer networking are mainly introduced.

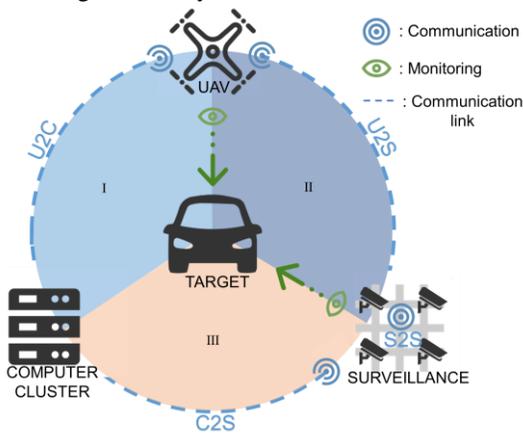

Figure 1. Architecture of UAV-aided urban target tracking system. I:case1 means normally tracking. II+III: case 2 means the target is about to be lost. III: case 3 means the target is completely lost.

### A. System scenarios

As a reliable and effective tracking system, once the target is lost, it can locate the target again in a short time by itself. In the UAV-aided urban target tracking system, response to the tracking task should be in real time. However, there always exist large communication and computing delays in processing a tracking task, which may cause the UAV-aided urban target tracking system to lose the target. Therefore the tracking scenarios was divided into three cases in order to improve the reliability and effectiveness of the tracking system.

1) *Case 1:Offloading the video processing task to the ground edge server to recognize the target in a near real-time manner.* Considering the limited energy of UAV, the video clips collected by UAV are sent to the ground nearby edge servers for analysis and processing. And then calculation results are transmitted to the UAV to guide its flight path to continue tracking.
2) *Case 2: Activating the ground cameras to relay tracking.* Once the target is about to be lost in the UAV's vision, the ground cameras are activated to survey the surrounding and then offload the video processing task to the ground edge server based on the strategy of case 1.
3) *Case 3: Offloading videos processing tasks from ground cameras to ground edge nodes to locate the lost target again in a short time.* It is the worst case that videos tasks captured by surveillance cameras need to be offloaded into nearby edge servers or assigned to edge mobiles. However, the offloading strategy in case 3 is different from that of case 1, that is, there are multiple independent and geographic-distributed camera videos that the processing tasks need to process at a given time in case 3 but in case 1, there is a UAV video streaming that needs to be offloaded to the ground edge server. Indeed, two different offloading strategies were designed according to the distribution between edge servers and edge mobile terminals on the ground.

### B. System Components

1) *UAV*: The UAV is used to collect and transmit the video clips of a suspected target when tracking. It is equipped with a camera, position sensors (a global positioning system, an acceleration transducer and gyroscope sensors), embedded micro-processors, and communication modules, such as LTE or Wi-Fi module. The UAV can capture some video clips of the target with the camera, and then it can communicate with the ground servers through LTE or Wi-Fi according to their distance. As for the flight path, the UAV can implement motion control through onboard sensing and processing or the instructions of the ground server. However, when the target is about to be lost in the UAV's vision, the UAV is used as a correspondent and it activates some ground surveillance cameras according to their own location. Once the target is found by the ground camera again, the UAV can re-plan its own route according to the target's motion and speed.
2) *The ground surveillance cameras*: The ground surveillance cameras are abundant, geographically distributed, and constitute more than one cable ground video surveillance network, such as road monitoring network, security monitoring network, violation monitoring network, or speed monitoring network. However, their computing capabilities are unable to support the real-time tracking task. Therefore, in order to keep up with the target effectively, ground surveillance cameras should be activated selectively according to their geographical location and road conditions, which can be obtained from GPS (global positioning system) and the aerial UAV. Each surveillance camera is equipped with storage, communication and micro-processing modules, which enable data transmissions among the ground computing cluster and the UAV. The transmission between cameras and the computing cluster can facilitate the cooperation of tracking computing task. In addition, ground surveillance cameras can also communicate with each other to cover all suspicious areas.
3) *The ground computing cluster*: The ground computing cluster is composed of ground edge servers, mobile phones, laptops and so on. Each device in the cluster has enough energy and some surplus computing power. If the cluster receives the UAV's video, it is necessary to choose one server to compute the tracking task according to the distance between the UAV and the servers and the computing power required by the task. Supposing that the cluster received multiple videos from the ground surveillance cameras at a time, it should make a decision of offloading or assignment of the computing task. The computing cluster is mainly responsible for the data analysis and processing, the scheduling of computing tasks and edge devices, and bridging between the UAV and the ground surveillance system.

*C. The three-layer networking*

As depicted in Figure 1, there exist three kinds of communication links in the UAV-aided urban target tracking system, including U2S/U2C links (among UAV and ground surveillance cameras or the computing cluster), S2S links (among ground surveillance cameras), and S2C links (among ground surveillance cameras and computing cluster), respectively. Due to the flexibility of the UAV, it can collect some videos about the target where the ground cameras cannot be deployed. On the other hand, there are abundant computing devices around the surveillance cameras. So, combining with the flexibility of the UAV and the richness of edge devices, the effectiveness and reliability of the ground surveillance system can be improved. The system can be divided into three layers networking. The first layer is the UAV, where the UAV collects video data about the target and transfers to the ground edge server. The second layer is the ground cameras network, where it helps the UAV to relay tracking when the UAV is about to lose the target, communicates with each other for full coverage of a suspicious area, and requests the nearby edge server or terminals to compute tracking tasks. The last layer is the ground computing cluster, where all the edge devices in the cluster are responsible for the computing task and transmit the results to the UAV for further guideline. The details are as follows.

1) *U2S/U2C links:* The UAV communicate with the ground surveillance cameras network or the ground computing cluster through LTE. It can provide a high speed interaction among the UAV and cameras or terminals, which increases the capacity and speed using a different radio interface together with core network improvements. However, the transmission data have some information about the target, such as some pictures of the target, present geographic information, and recognition mode of the target. The first-second layer interaction is usually used to help the UAV to relay tracking, while the first-third layer interaction is used to help the UAV compute the tracking task when it is tracking the target normally.
2) *S2S links:* The ground surveillance cameras in the network can communicate with each other by LAN (Local Area Network). Compared with the wireless communication, the wire communication is more reliable and efficient. Therefore, once some cameras are activated by the UAV, they can broadcast in the wire network with minimum propagation time to achieve the full coverage of the target area in a way that is not redundant.
3) *S2C links*: The communication method between the ground surveillance cameras and the ground computing cluster is Wi-Fi, because its high data rates are supported in 802.11 protocols, and each cluster that is nearby the cameras network can easily access the LAN. This layer is especially important when the UAV has lost the target completely in other to locate the target again. So the data are usually some videos that were obtained during the surveillance time, and are offloaded to the edge terminals to find the target again.

IV. STRATEGIES AND MECHANISM FOR URBAN TARGET TRACKING SYSTEM

Four strategies were designed to improve the reliability and effectiveness of the urban target tracking system among three cases. Figure 2 shows the system prototype. The UAV activates the ground surveillance camera to collect the video information of the target, and then the camera requests the nearby computing node to process the video and recognize the target and the result further guides the UAV. DJI UAV wizard 3 was used, and it was tested for performing single vehicle tracking using BOOSTING method from Open Source Computer Vision (OpenCV). In this section, the details about these strategies are shown to improve the reliability and real-time of the system and some preliminary evaluation results.

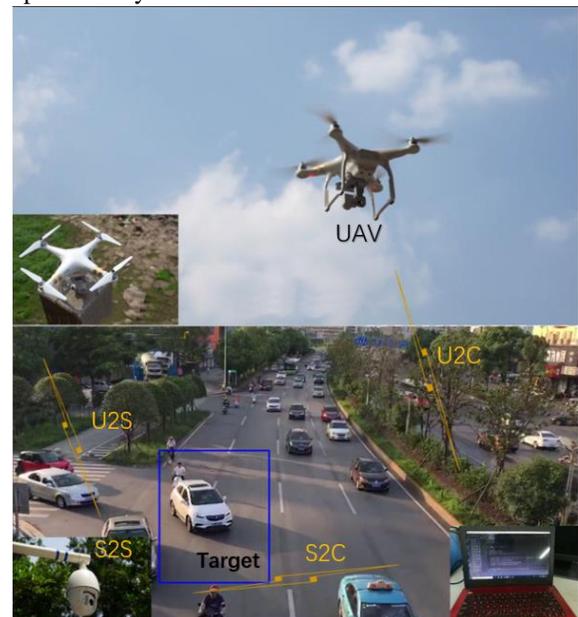

Figure 2. The system prototype in case 2.

1) *Mobile Resource Aware (MRA) stream processing scheduling:* If there is more than one edge server on the ground in case 1, a strategy should be designed to choose the most appropriate server to complete the video processing task so that the task can be completted as soon as possible. The best match of the CPU resource and memory resource between the processing task and the ground edge servers are the factors considered. However, the available edge servers set changes with the UAV's movement, and the task is non-independent in each slot. MRA scheduling algorithm [7] was implemented in Storm through a custom scheduler and the average processing time of MRA was evaluated in an emulation mobile edge environment. As shown in Figure 3, the average processing time of tuples with MRA is lower than the default scheduling algorithm and RAS, because the default one does not consider the communication between components, while the MRA does not consider the mobility of the UAV. So MRA is optimized for the mobile tracking environment.

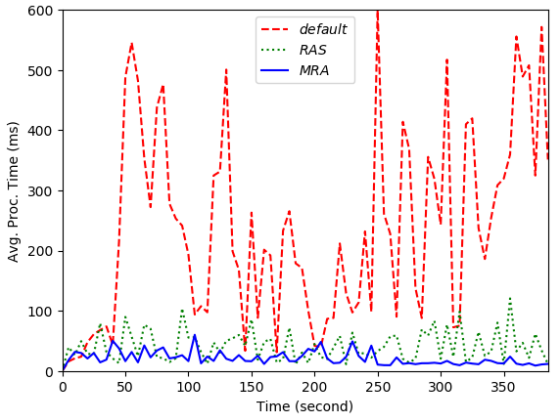

Figure 3. The average processing time of tuples. X-axis: the UAV's flight time. Y-axis: the task processing time.

2) *Activation strategy for relaying tracking(ASRT)*: In case 2, the UAV cannot locate the target temporarily because the target was blocked by obstacles. In order to continue to track the target, the UAV need to cooperate with the ground surveillance cameras and ground edge servers. However, the limited energy capability and delay-strict made the UAV to be unable to activate all the surveillance cameras to relay tracking the target. Therefore, an ASRT algorithm was designed to improve the real-time and effectiveness in the system. According to the task request delay and the response delay, a ring activation region can be gotten, which is made up of a ground wire surveillance network. Every camera in the network can communicate with neighbor nodes if there are communication links among them, so the UAV can activate some key cameras and other cameras are activated by these key cameras. All the possible activation sequences can be run to minimize the total activation time, so that these key cameras can be located in the ground surveillance network. As shown in Figure 4, the communication delay can be further shortened and the duration of the UAV can be prolonged.

3)

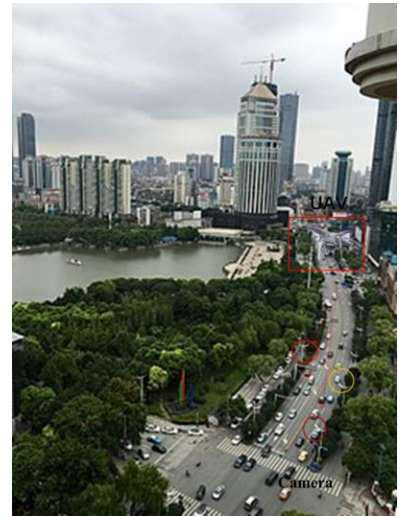

Figure 4. The model of activating some certain surveillance cameras. **Red rectangle box**: UAV
**Red ellipse box**: activated cameras
**Yellow ellipse box**: non-activated camera

4) *Strategy QoE-driven computation offloading*: In case 3, the UAV completely loses its target, the video from the ground surveillance cameras that were deployed near the target vanishing point need to be analyzed. However, there are higher delay and limited bandwidth using traditional methodof uploading all the video to the cloud. A new method of offloading video processing tasks to the edge servers is proposed if there are some available edge servers. Offloading the processing task whether to the terminal node or to the edge server depends on the video data size and the required CPU cycles [8]. The execution time and the corresponding energy consumption for each video processing task offloading to terminal node or edge server can be calculated. And then by comparing the total latency time and energy consumption on the terminal node or on the edge server, the pattern of smaller cost of offloading scheme of video processing tasks can be obtained. Compared to benchmark schemes, it has been demonstrated that the proposed offloading scheme improved the performance on latency time and energy consumption, that is better QoE. As shown in Figure 5.

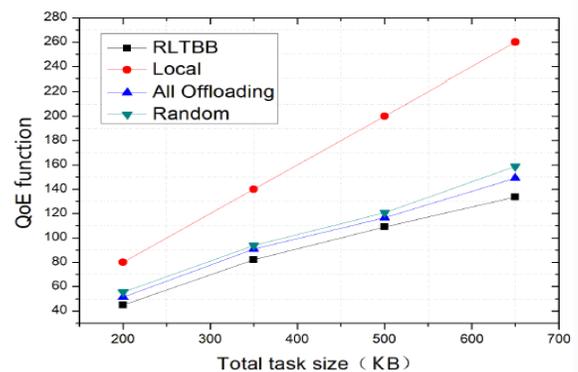

Figure 5. Performance comparison on QoE function.

5) *Cloudlet Assisted Cooperative Task Assignment(CACAT) Strategy*: In case 3, if there are rich available terminals near

the vanishing point, a different method of offloading the analysis task to the terminals should be proposed because of shorter communication delay. Each video processing task has its own workload and time requirement due to different video data size. When the cloudlet receives the information, it broadcasts the request to the edge nodes nearby and organizes them into a collaboration cluster to carry out the task and each node need to give feedback of its status, such as its capacity, cost, and available time. On the other hand, the allocator or the workload assignment module in the cloudlet divides a task into a collection of subtasks (e.g., divide a video into different sets of frames in a video processing application) to match edge nodes, and then assigns them to the edge nodes in the collaboration cluster in multiple rounds according to a Prediction-based Assignment Optimization (PA-opt) algorithm [9]. As soon as the collaboration cluster completes the task, the cloudlet returns the result to the UAV which submits the task request. As shown in Figure 6, the proposed algorithm has better performance on the competition ratio $C_{rA}$, which is defined as the ratio of the total costs and the offline optimization solution.

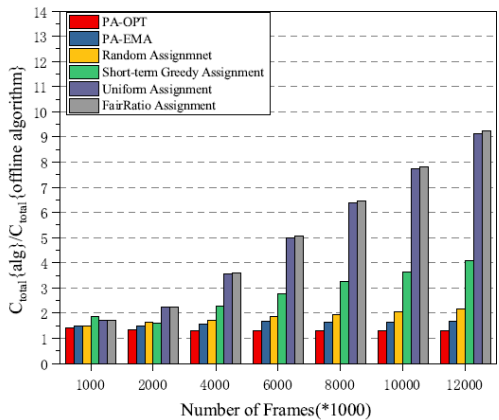

Figure 6. Experiments with different assignment algorithms.

## V. CONCLUSION AND DISCUSSION

In this article, state-of-art applications of incorporating UAVs into ground networks have been reviewed. Due to the complexity of tracking task, an aerial-ground cooperative surveillance system, which is composed of an UAV, a ground surveillance cameras network and an edge computing cluster on the ground have been proposed. Some effective algorithms were designed to deal with three important challenges in urban target tracking, including 1) offloading UAV's video processing task to the ground edge server by MRA stream processing scheduling in case1. 2) Effectively Activating ground surveillance cameras to relay tracking by ASRT in case2. 3) Offloading video processing tasks from ground cameras to ground edge nodes to locate the lost target again in a short time in case 3. Relative effective algorithms were designed to deal with the possibility of losing the target of UAV, and the simulation results have demonstrated that they can reduce the computing latency and improve the reliability of the tracking system. In the future, effort will be concentrated toward deploying the tracking system with the above strategies. Multiple UAV will also be designed to study the real-time and reliability of tracking system. This study is expected to provide a new solution to facilitate future applications in smart city for public safety system.